\documentclass[showpacs,superscriptaddress,preprintnumbers,amsmath,amssymb,onecolumn]{revtex4}

\usepackage{graphicx}
\usepackage{dcolumn}
\usepackage{bm}

\usepackage{color}

\begin{document}

\title{Nonlinear Conductivity of a Holographic Superconductor Under  Constant Electric Field}

\author{Hua Bi Zeng}\email{zenghbi@gmail.com}
\affiliation{College of Physics Science and Technology, Yangzhou University, Jiangsu 225009, China}
\affiliation{Department of Physics, National Central University, Chungli 32001, Taiwan}
\affiliation{School of Mathematics and Physics, Bohai University, JinZhou 121000, China}

\author{Yu Tian}\email{ytian@ucas.ac.cn}
\affiliation{School of Physics, University of Chinese Academy of Sciences, Beijing 100049, China}
\affiliation{Shanghai Key Laboratory of High Temperature Superconductors, Shanghai 200444, China}

\author{Zhe Yong Fan}\email{brucenju@gmail.com}
\affiliation{COMP Centre of Excellence, Department of Applied Physics, Aalto University, Helsinki, Finland}

\author{Chiang-Mei Chen}\email{cmchen@phy.ncu.edu.tw}
\affiliation{Department of Physics, National Central University, Chungli 32001, Taiwan}

\begin{abstract}
The dynamics of a two-dimensional superconductor under a constant electric field $E$ is studied by using the gauge/gravity correspondence. The pair breaking current induced by $E$ first increases to a peak value and then decreases to a constant value at late time, where the superconducting gap goes to zero, corresponding to a normal conducting phase. The peak value of the current is found to increase linearly with respect to the electric field. Moreover, the nonlinear conductivity, defined as an average of the conductivity in the superconducting phase, scales as $\sim E^{-2/3}$ when the system is close to the critical temperature $T_c$, which agrees with predictions from solving the time dependent Ginzburg-Landau equation. Away from $T_c$, the
$E^{-2/3}$ scaling of the conductivity still holds when $E$ is large.
\end{abstract}


\maketitle
\pagebreak

\section{Introduction}

AdS/CFT correspondence, or holography, provides a remarkable method to investigate strongly coupled quantum field theory problems by solving the classical equations of motion of the gravity dual~\cite{Maldacena:1997re, Gubser:1998bc, Witten:1998qj, Aharony:1999ti}. The partition function of a quantum many-body system can be computed by the path integral approach in quantum field theory. Thus, due to the holographic principle, the AdS/CFT correspondence can be used to study a quantum many-body problem by computing its partition function from the gravity dual theory~\cite{Zaanen:2015oix,Ammon2015,Hartnoll:2009sz, McGreevy:2009xe, Sachdev:2010ch, Green:2013fqa}. For example, the holographic superconductor model, as proposed in Refs.~\cite{Gubser:2008px, Hartnoll:2008vx}, has been demonstrated to be quite successful in describing the superconducting equilibrium state phase transitions and charge transport properties in the linear response regime~\cite{Cai:2015cya, Herzog:2009xv}. Moreover, holography can also be used to obtain the dynamics of a quantum many-body system  by solving the classical time-evolution equations of its gravity dual, regardless of the system being close to or far from equilibrium. For example, holography has been applied to study the dynamics of the superconducting gap far from equilibrium, in spatially homogeneous ~\cite{Murata:2010dx, Bhaseen:2012gg, Gao:2012aw, Bai:2014tla, Li:2013fhw, Basu:2011ft, Basu:2012gg} or inhomogeneous~\cite{Sonner:2014tca, Garcia-Garcia:2013rha, Adams:2012pj, Du:2014lwa} configuration.

Nonlinear transport appears when the external electric field $E$ cannot be treated as a perturbation, where the linear response theory breaks down. In this case, one can still use the AdS/CFT correspondence to study the dynamics of the induced current, which provides a way to compute the nonlinear conductivity. For example, holography has been used to compute the non-superconducting steady current driven by a constant or oscillating electric field~\cite{Karch:2010kt, Sonner:2012if, Kundu:2013eba,Horowitz:2013mia}, where a constant nonlinear conductivity independent of the strength and frequency of the electric field has been found in a two-dimensional field theory. In Ref.~\cite{Zeng:2016api}, we have studied the nonlinear correction to the AC conductivity of a two-dimensional holographic superconductor, where an $E^2$ scaling nonlinear conductivity in the superconducting phase has been found. In this paper, we extend the study of nonlinear conductivity to consider a constant electric field applied to the holographic superconductor. We find that the induced current attains a peak value proportional to $E$ and the nonlinear conductivity scales as $\sim E^{-2/3}$ in the limit of large $E$, which agrees well with predictions from solving the time-dependent Ginzburg-Landau equation~\cite{Kajimura, Dorsey}.

This paper is organized as follows. Section~2 introduces our model. Section~3 discusses the dynamics of the current and the superconducting gap. Section~4 discusses the nonlinear conductivity and section~5 concludes.

\section{Gravity dual of superconductor under constant electric field}

The minimal version of the holographic superconductor model in the $(d+1)$-dimensional anti-de Sitter space AdS$_{d+1}$ is a Maxwell-Higgs model with the action 
\begin{equation}
S = \int d^{d+1}x \sqrt{-g} \left( - \frac{1}{4} F_{\mu\nu} F^{\mu\nu} - |D \Psi|^2 - m^2 |\Psi|^2 \right).
\end{equation}
The metric of the AdS$_{d+1}$ black brane in the Eddington coordinates is
\begin{equation}
ds^2 = \frac{L^2}{z^2} \left( -f(z) dt^2 - 2 dt dz + dx^i dx_i \right),
\end{equation}
where $f(z) = 1 - z^d$. Choosing $L = 1$ and $m^2=-(d^2-1)/4$ leads to $\sqrt{d^2 + 4 m^2 L^2} = 1$. The nonzero fields in our setup are $A_t(t, z)$, $A_x(t, z)$, and $\Psi(t, z) = \Phi(t, z) z^{(d-1)/2}$, and one can derive the equations of motion,
\begin{eqnarray}
\partial_t \left( z^{3-d} \partial_z A_t \right) + 2 A_t \Phi^2 - i f \left( \Phi^* \partial_z \Phi - \Phi \partial_z \Phi^* \right) + i ( \Phi^* \partial_t \Phi - \Phi \partial_t \Phi^* ) &=& 0,
\\
\partial_t \left( z^{3-d} \partial_z A_x \right) + \partial_z \left( z^{3-d} \partial_t A_x \right) - \partial_z \left( z^{3-d} f \partial_z A_x \right) + 2 A_x \Phi^2 &=& 0,
\\
\partial_t \partial_z \Phi - i A_t \partial_z \Phi + \frac12 \Bigl[  \partial_z (f \partial_z \Phi) + i \partial_z A_t \Phi - A_x^2 \Phi \qquad &&
\nonumber\\
- \frac{(d^2 - 1)(f - 1) - 2(d - 1) z \partial_z f}{4 z^2} \Phi \Bigr] &=& 0, \label{eqPhi}
\end{eqnarray}
along with a constraint 
\begin{equation}
\partial_z \left( z^{3-d} \partial_z A_t \right) - i ( \Phi^* \partial_z \Phi - \Phi \partial_z \Phi^* ) = 0.
\end{equation}
The equation of motion of $\Phi$ can also be expressed as
\begin{equation}
\partial_t \partial_z \Phi - i A_t \partial_z \Phi + \frac12 \Bigl[  \partial_z (f \partial_z \Phi) + i \partial_z A_t \Phi - A_x^2 \Phi - \left( \frac{d-1}2 \right)^2 z^{d-2} \Phi \Bigr] = 0.
\end{equation}

In the above equations, a gauge in which $A_z = 0$ is chosen. To be specific, we focus on the case of $d = 3$, which corresponds to a two-dimensional superconductor on the boundary. This should be relevant to the experiment of Kajimura \textit{et al.}~\cite{Kajimura}, where the nonlinear conductivity of aluminum films below the critical temperature under a constant electric field has been measured.

The asymptotic behaviors of the scalar field and gauge fields 
\begin{equation}
\Phi(z \rightarrow 0) = \Phi^{(1)} + \Phi^{(2)} z,
\end{equation}
and
\begin{equation}
A_\nu(z \rightarrow 0) = a_{\nu} - b_{\nu} z.
\end{equation}
We employ the standard quantization of the scalar field, where $\Phi^{(1)}$ is regarded as the source and $\Phi^{(2)}$ is identified as the expectation value $\langle O_n \rangle$. We expect to get similar results in the alternative quantization according to previous works. The boundary condition of $\Phi = 0$ at the boundary $z = 0$ is imposed to guarantee the spontaneous symmetry breaking. In the Eddington coordinates, $a_{\nu}$ are the gauge fields on the boundary, and the current along the $x$ direction of the boundary field theory is given as~\cite{Li:2013fhw}
\begin{equation}
\quad J_x =  b_x + \partial_t a_x.
\end{equation}
The chemical potential $\mu$ and charge density $\rho$ are connected to the time component of the gauge field:
\begin{equation}
\mu = a_t,  \qquad \rho = b_t + \partial_t a_t.
\end{equation}
At $t = 0$, we prepare an equilibrium superconducting state at a fixed temperature, which can be obtained by solving the time-independent version of the above equations for $\Phi$ and $A_t$ by setting $A_x = 0$.
The scalar field here is a complex field, and its imaginary part is set to zero at the horizon to fix a degree of freedom.



\begin{figure}
\includegraphics[trim=1cm 0cm 1cm 0.5cm, clip=true,scale=0.65]{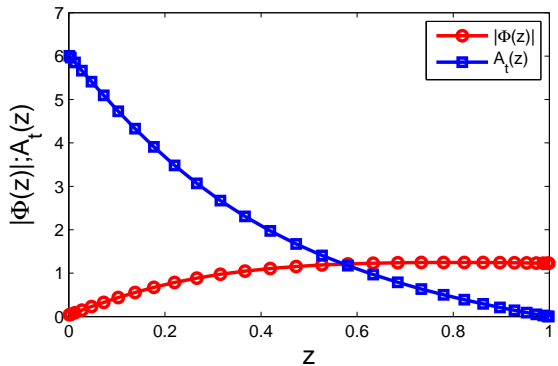}
\caption{The initial equilibrium configuration of $|\Phi(z)|$ and $A_t(z)$ at $T = 0.68 T_c$. Here, $z = 1$ is the horizon and $z = 0$ is the boundary at infinite.} \label{figure:1}
\end{figure}

There is a critical value $\mu_c$ of the chemical potential, above which the system enters a $U(1)$ gauge symmetry breaking state with non-zero $\Phi$. The dimensionless temperature $T/T_c$, $T_c$ being the critical temperature, is defined as $\mu_c/\mu$, since the temperature $T$ is proportion to $1/\mu$. Figure~\ref{figure:1} shows a typical equilibrium state at $t = 0$, where $\mu = 6$ and $T = 0.68 \, T_c$. The phase of the scalar field is also $z$-dependent, which can be transformed to $A_z$ according to the gauge invariance.  Therefore, we can prepare the initial state by setting both the phase of the scalar field and $A_z$ to zero in our choice of gauge. After preparing the initial state, we apply a constant electric field $E$ at the boundary, which is achieved by setting
\begin{equation}
a_x = E t.
\end{equation}

The dynamics of the system under the constant electric field can be obtained by solving the time-dependent equations of motion. We numerically solve these equations by combining a pseudo-spectral method and the Runge-Kutta method. The bulk fields are first expanded using Chebyshev polynomials in the $z$ direction, and then plugged into the partial differential equations. This results in a set of ordinary differential equations, which can then be solved by the fourth-order Runge-Kutta method~\cite{Li:2013fhw, Zhang:2016coy}.

\section{Dynamics of superconducting gap and induced current}

The electric field will drive the charge carrier to flow and thus induce a current. An oscillating electric field will induce an oscillating current with the same frequency, and the current lags behind the field, resulting in a complex AC conductivity. Nonlinear correction to the AC conductivity occurs when the induced current affects the superconducting gap dramatically. In Ref.~\cite{Zeng:2016api}, we have studied the current dynamics in the nonlinear regime and an $E^2$ scaling of the nonlinear AC conductivity has been predicted.

\begin{figure}
\includegraphics[trim=1cm 0cm 1cm 0.5cm, clip=true,scale=0.65]{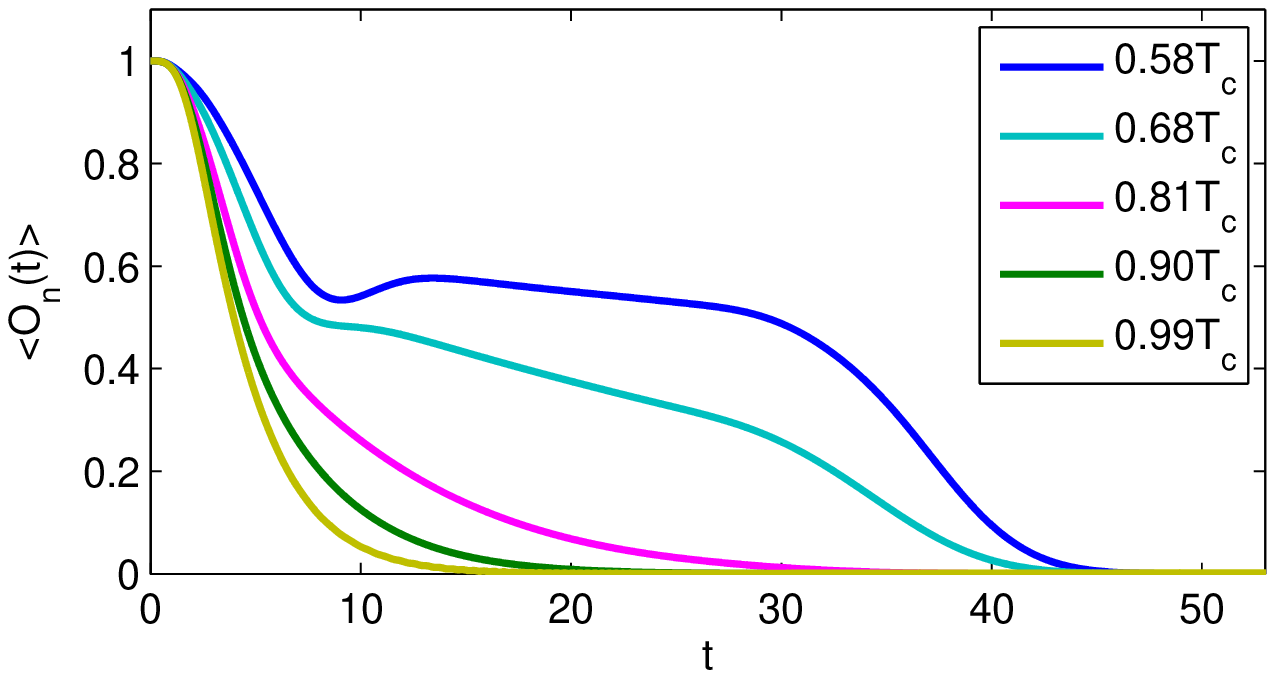}
\includegraphics[trim=1cm 1cm 1cm 0.5cm, clip=true,scale=0.65]{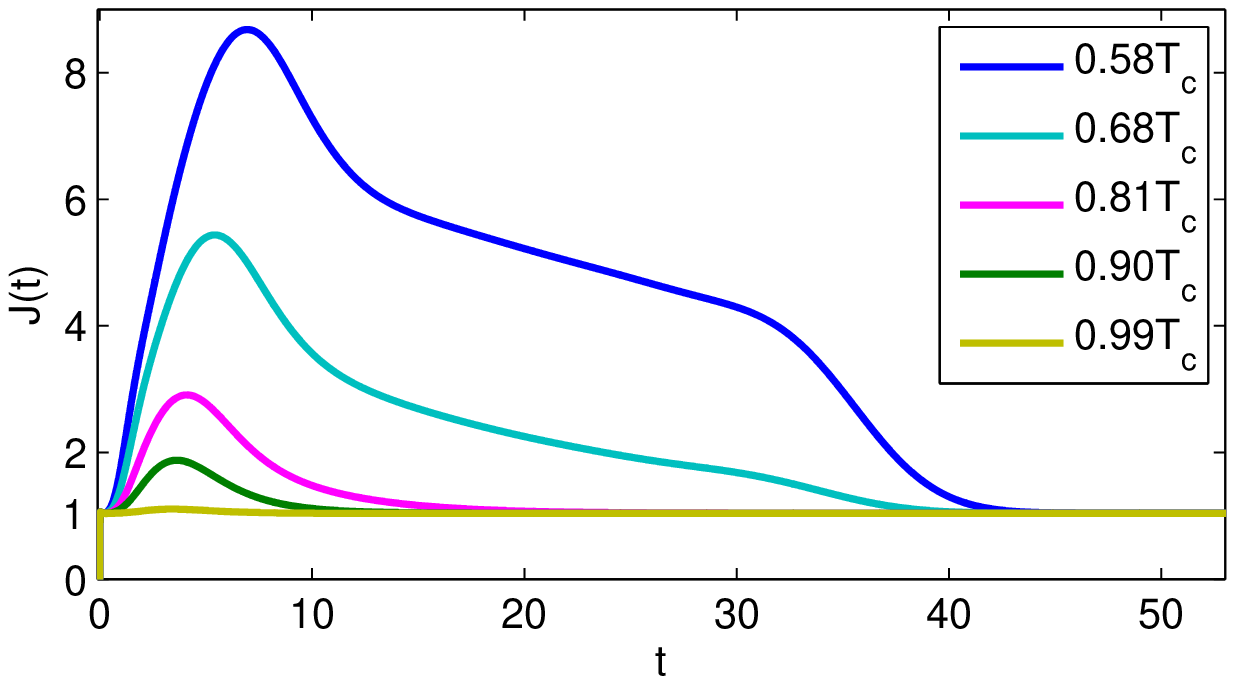}
\caption{The superconducting gap $\langle O_n \rangle$ (upper panel) and the induced current $J$ (lower panel) as a function of time $t$ at different temperatures. The electric filed here is $E = 1$. }
\label{figure:2}
\end{figure}

\begin{figure}
\includegraphics[trim=1cm 0cm 1cm 0.5cm, clip=true,scale=0.65]{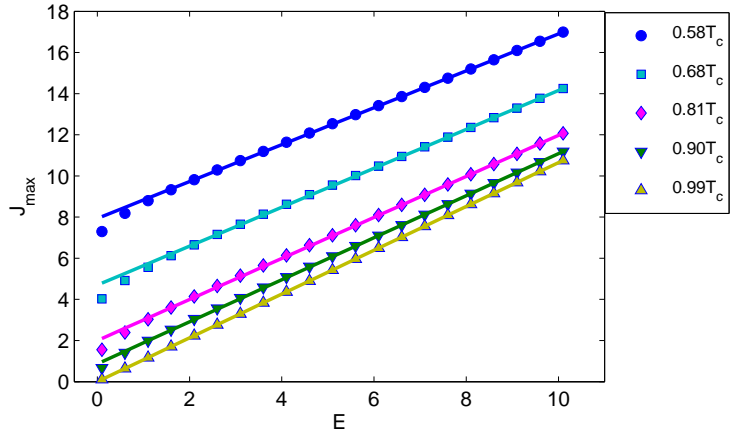}
\caption{The maximum current $J_\mathrm{max}(E)$ as a function of the electric field $E$ at different temperatures.}
\label{figure:3}
\end{figure}

In contrast to the case of oscillating electric field, a constant electric field will keep enhancing the current up to a critical value, where the kinetic energy of the cooper pair is comparable to the superconducting gap, which can destroy the superconducting state. According to the London equation, $\partial_t J(t) \propto E$, the current $J(t)$ will increase linearly with time initially and then drops to a value of the normal state, $\sigma_n$, when the kinetic energy of the cooper pair exceeds the superconducting gap. More quantitative results can be obtained by numerically solving the time-dependent equations. Figure~\ref{figure:2} shows the results with $E = 1$ at different temperatures below $T_c$. The superconducting gap $\langle O_n \rangle$  decreases rapidly with increasing time and vanishes at large time, where the system enters the normal state. When $T$ is close to $T_c$,  $\langle O_n(t) \rangle$ decays exponential with time. At low temperatures,  $\langle O_n(t) \rangle$ behaves in a more complicated way at intermediate time, but still decays exponentially at large time. On the other hand, the current $J(t)$ first reaches a peak value $J_\mathrm{max}$ and then deceases to a constant value of $J = E$ at large time, where the superconducting state is completely destroyed. When the system enter the normal state, we have a constant linear conductivity of $\sigma=J/E=1$, which agrees with previous works~\cite{Karch:2010kt, Sonner:2012if, Horowitz:2013mia, Zeng:2016api}. When $T$ is closed to $T_c$, $J(t) - 1$ behaves as $\sim t e^{-t}$. Similar results have been obtained for other values of $E$. At a fixed temperature, $J_\mathrm{max}$ scales almost linearly with respect to $E$, as shown in Fig.~\ref{figure:3}. Quantitatively, we have $J_\mathrm{max}(T) \approx k E + J_0(T)$, where the proportional constant is determined to be $k \approx 0.9$ and $J_0$ can be regarded as the critical superconducting current under zero electric field. We find that $J_0(T)$ scales as $\sim (T_c - T)^{3/2}$ in the vicinity of the critical temperature, which is in consistent with previous works on holographic superconductor~\cite{Zeng:2010fs, Arean:2010xd,Arean:2010zw, Zeng:2012xy} as well as the Ginzburg-Landau theory.

\section{Nonlinear  conductivity}



\begin{figure}
\includegraphics[trim=1.0cm 0cm 1cm 0.5cm, clip=true,scale=0.65]{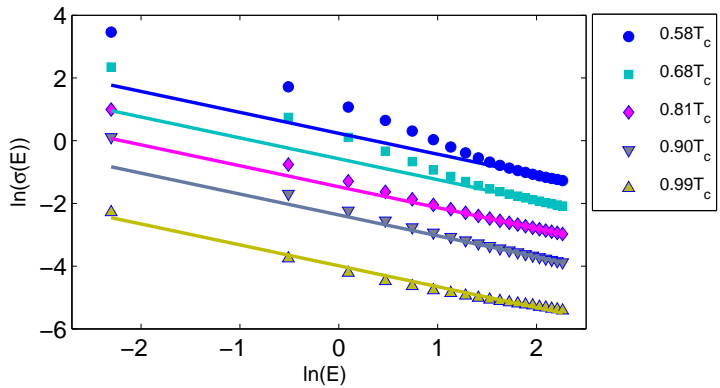}
\caption{The nonlinear DC conductivity $\sigma(E)$ as a function of the electric field $E$ at different temperatures. The lines are fits to the data in the large $E$ limit using Eq.~(\ref{equation:sigma_fit}).}
\label{figure:5}
\end{figure}

Following Refs.~\cite{Kajimura, Dorsey}, a nonlinear DC conductivity $\sigma(E)$ in the superconducting phase under a constant $E$ can be defined by
\begin{equation}
\sigma(E) = \frac{\langle J \rangle}{E} = \frac{\int_0^{t_\mathrm{max}} J(t) dt}{t_\mathrm{max} E},
\end{equation}
where $t_\mathrm{max}$ is the time at which the system enters the normal state. Numerically, $t_\mathrm{max}$ is determined as the time at which the order parameter decays to a small value of $10^{-3}$. Figure~\ref{figure:5} shows the calculated nonlinear DC conductivities at different temperatures as a function of the electric field. The lines are power-law fittings to the data in the limit of large $E$:
\begin{equation}
\label{equation:sigma_fit}
\sigma(E) \propto E^{\alpha}.
\end{equation}
It is found that a value of $\alpha=-2/3$ results in an optimal match between Eq.~(\ref{equation:sigma_fit}) and the numerical data. In the limit of small $E$, the deviation of the data from Eq.~(\ref{equation:sigma_fit}) becomes larger for lower temperature. However, in the limit of large $E$,  Eq.~(\ref{equation:sigma_fit}) still holds even for lower temperatures. Interestingly, the scaling behavior of the nonlinear DC conductivity in our model agrees with that predicted by the scaling theory for a $d$-dimensional superconductor~\cite{Dorsey}:
\begin{equation} \label{sigma}
 \sigma(E) \propto E^{-(2 - d + z)/(z + 1)},
\end{equation}
as the dynamic exponent in our holographic superconductor model is $z = 2$~\cite{Maeda:2009wv, Jensen:2011af}. The scaling can be obtained quickly by a dimension analysis, the electric field, in a gauge where the scalar potential is zero, is simply $\mathbf{E} = \partial \mathbf{A} / \partial t$, so that the electric field has the dimension of inverse length times inverse time. Since the characteristic unit of time is the order parameter relaxation time $\tau \sim \xi^z$, and the length scale is $\xi$, we see that the characteristic scale for the electric field is $\xi^{-(1 + z)}$. The scaling form for the nonlinear conductivity is thus
\begin{equation}
 \sigma(E) \propto \xi^{2 - d + z} \sum (\xi^{1 + z} E),
\end{equation}
where $\sum$ is the universal scaling functions around $T_c$. At $T = T_c$, the correlation length diverges, and the nonlinear conductivity takes the form Eq.~(\ref{sigma}). This $E^{-2/3}$ scaling has been observed in the measured nonlinear conductivity in superconducting aluminum films~\cite{Kajimura} close to $T_c$, which can be well understood in terms of the phenomenological time-dependent  Ginzburg-Landau theory~\cite{Kajimura, Dorsey}. The survival of the $E^{-2/3}$ scaling away from $T_c$ observed in our data, however, is a new prediction by the non-perturbative holography theory.






\section{Conclusions}


In conclusion, we have studied nonlinear transport in a two-dimensional superconductor using the gauge/gravity correspondence. Near the critical point we have found that the nonlinear conductivity scales as $\sim E^{-2/3}$ with respect to the electric field $E$, which agrees with both experiments~\cite{Kajimura} and results obtained from solving the time dependent Ginzburg-Landau equation~\cite{Kajimura, Dorsey}.
Away from $T_c$, the $E^{-2/3}$ scaling also appears when the external field $E$ is large. We have also predicted that the maximum current induced by the external constant electric field increases linearly with $E$. New experiments are needed to test our predictions regarding the $\sim E^{-2/3}$ scaling of the nonlinear conductivity away from $T_c$ and the $\sim E$ scaling of the maximal current in a two-dimensional superconductor.

\section*{Acknowledgements}
We thank Hai Qing Zhang for sharing his code and experiences. HBZ and ZF are supported by the National Natural Science Foundation of China (under Grant No. 11675140 and No. 11404033). Y.T. is partially supported by NSFC with Grant No.11475179 and the Opening Project of Shanghai Key Laboratory of High Temperature Superconductors (14DZ2260700). CMC is supported by the Ministry of Science and Technology of Taiwan under the grants MOST 102-2112-M-008-015-MY3 and MOST 105-2112-M-008-018.

\end{document}